\documentclass[
    ,final            
  ]
  {aipproc}

\layoutstyle{6x9}

\begin{document}

\title{The LSST Data Mining Research Agenda}


\author{K. Borne}{
  address={Computational and Data Sciences, George Mason University, MS 6A2, Fairfax, VA 22030}
}

\author{J. Becla}{
  address={Stanford Linear Accelerator Center, Stanford University, Stanford, CA 94309}
}

\author{I. Davidson}{
  address={Department of Computer Science, University of California, Davis, CA 95616}
}

\author{A. Szalay}{
  address={Department of Physics \& Astronomy, Johns Hopkins University, Baltimore, MD 21228}
}

\author{J. A. Tyson}{
  address={Physics Department, University of California, Davis, CA 95616}
}

\begin{abstract}
We describe features of the LSST science database that are amenable to scientific data mining, object classification, outlier identification, anomaly detection, image quality assurance, and survey science validation. The data mining research agenda includes: scalability (at petabytes scales) of existing machine learning and data mining algorithms; development of grid-enabled parallel data mining algorithms; designing a robust system for brokering classifications from the LSST event pipeline (which may produce 10,000 or more event alerts per night); multi-resolution methods for exploration of petascale databases; indexing of multi-attribute multi-dimensional astronomical databases (beyond spatial indexing) for rapid querying of petabyte databases; and more. 
\end{abstract}

\maketitle


\section{Data-Intensive Astronomy and The LSST Sky Survey}

The development of models to describe and understand 
scientific phenomena has historically proceeded at a 
pace driven by new data.  The more we know, the more 
we are driven to tweak or to revolutionize our models, 
thereby advancing our scientific understanding. 
This data-driven modeling and discovery linkage has 
entered a new paradigm \cite{mahoot08}.  The acquisition of 
scientific data in all disciplines is now accelerating 
and causing a nearly insurmountable data avalanche \cite{bell05}. 
In astronomy in particular, rapid advances in three 
technology areas (telescopes, detectors, and computation) 
have continued unabated \cite{gray2005} --  
all of these advances lead to 
more and more data \cite{becla06}. With this accelerated advance in 
data generation capabilities, we will require novel, 
increasingly automated, and increasingly more effective 
scientific knowledge discovery systems \cite{borne06}. 

Astronomers have been doing
data mining for centuries: {\it{``the data are mine,
and you can't have them!''}}.  Seriously, astronomers
are trained as data miners, because we are trained to: 
(a)~characterize the known 
({\it{i.e.}}, unsupervised learning, clustering);
(b)~assign the new
({\it{i.e.}}, supervised learning, classification);
and 
(c)~discover the unknown
({\it{i.e.}}, semi-supervised learning, outlier detection)
\cite{borne01A, borne01B, borne2009}.
These skills are more critical than ever since 
astronomy is now a data-intensive science, and it will
become even more data-intensive in the coming decade
\cite{becla06, brunner2001, szalay2002}.
New surveys may produce hundreds of terabytes (TB) up to 
100 (or more) petabytes (PB) both in the image data archive 
and in the object catalogs (databases). 
Discovering the ensuing hidden wealth of new 
scientific knowledge will require more sophisticated 
algorithms and networks that discover, integrate, and 
learn from distributed petascale databases more effectively
\cite{gray2002}, \cite{longo2001}.

The problem therefore is this: astronomy researchers 
will soon (if not already) lose the ability to assimilate
or to keep up 
with any of these things: the data flood, the scientific 
discoveries buried within, the development of new models 
of those phenomena, and the resulting new data-driven 
follow-up observing strategies that are imposed on 
telescope facilities to collect new data needed to 
validate and augment new discoveries.


One of the most impressive astronomical sky surveys 
being planned for the next decade is the Large Synoptic 
Survey Telescope project (LSST at www.lsst.org) \cite{tyson04}.  
The three fundamental distinguishing astronomical attributes 
of the LSST project are: 

\begin{enumerate}

\item {\it{Repeated temporal measurements}} of all observable 
objects in the sky, corresponding to thousands of observations 
per each object over a 10-year period, expected to generate 
10,000-100,000 alerts each night to the astronomical research 
community that something has changed at that location on the sky: 
either the brightness or position of an object, or the 
serendipitous appearance of some totally new object;

\item {\it{Wide-angle imaging}} that will repeatedly cover 
most of the night sky within 3 to 4 nights (= tens of 
billions of objects); and 

\item {\it{Deep co-added images}} of each observable 
patch of sky (summed over 10 years: 2015-2025), 
reaching far fainter objects and to greater distance 
over more area of sky than other sky surveys \cite{strauss04}. 

\end{enumerate}

Compared to other astronomical sky surveys, the LSST survey 
will deliver time domain coverage for orders of magnitude more 
objects. It is envisioned that this project will 
produce $\sim$30 TB of data per each night of observation for 10 years.  
The final image archive will be greater than 60 PB (and
possibly much more), and the final LSST 
astronomical object catalog (object-attribute database) 
is expected to be $\sim$10-20 PB (or more).  Additional information
about the LSST survey and scientific program are
described by Ivezic et al. \cite{ivezic2008a} and provided
elsewhere in these proceedings \cite{ivezic2008b}.

Since it is anticipated that LSST will generate many 
thousands (probably tens of thousands) of new astronomical 
event alerts per night of observation, there is a critical 
need for innovative follow-up procedures.  These procedures 
necessarily must include modeling of the events -- 
to determine their classification, time-criticality, 
astronomical relevance, rarity, and the scientifically 
most productive set of follow-up measurements.  Rapid 
time-critical follow-up observations, with a wide range 
of time scales from seconds to days, are essential for 
proper identification, classification, characterization, 
analysis, interpretation, and understanding of nearly 
every astrophysical phenomenon ({\it{e.g.}}, supernovae, novae, 
accreting black holes, microquasars, gamma-ray bursts, 
gravitational microlensing events, extrasolar planetary 
transits across distant stars, new comets, incoming 
asteroids, trans-Neptunian objects, dwarf planets, 
optical transients, variable stars of all classes, 
and anything that goes ``bump in the night'') 
\cite{pacz2000, becker2008}.

\subsection{Petascale Mining of Large Astronomical Sky Surveys}
\label{s:lsst-dm}

LSST and similar large sky surveys have enormous potential 
to enable countless astronomical discoveries.  
Such discoveries will span the full spectrum of statistics: 
from rare one-in-a-billion (or one-in-a-trillion) 
type objects, to a complete statistical and astrophysical 
specification of a class of objects (based upon millions 
of instances of the class).  One of the key scientific 
requirements of these projects therefore is to learn 
rapidly from what they see. This means: 
(a)~to identify 
the serendipitous as well as the known; 
(b)~to identify rare events that our models say should be there; 
(c)~to identify 
new classes of objects that 
fall outside the bounds of model expectations; 
(d)~to find new attributes of known classes; 
(e)~to provide statistically robust tests of existing models; 
and (f)~to generate the vital inputs for new models.  
All of this requires integrating and mining all known data: 
to train classification models and to apply classification models.

LSST alone is likely to throw such data mining and knowledge 
discovery efforts into the petascale realm.  For example: 
astronomers currently discover a few hundred new supernovae 
per year. Since the beginning of human 
history, perhaps $\sim$10,000 supernovae have been recorded.  
Because the identification, classification, and analysis of 
supernovae enable fundamental (Dark Energy) science, 
it is imperative 
for astronomers to respond quickly to each new event with 
rapid follow-up observations in many measurement modes 
(light curves; spectroscopy; and images of the host galaxy
and its environment).  Historically, with $<$10 new supernovae 
being discovered each week, such follow-up has been feasible.  
But now, LSST promises to produce a list of 
{\it{1000 new supernovae each night}} for 
10 years \cite{strauss04}, which represent a small fraction of the 
total (10-100 thousand) alerts expected each night! 
Astronomers are faced with the enormous challenge of 
efficiently mining, correctly classifying, and 
intelligently prioritizing a staggering number of 
new events for follow-up observation each night for a decade.

The major features and contents of the LSST scientific database include: $>$100 database tables;
image metadata (675M rows);
source catalog (260B rows); 
object catalog (22B rows, with 200+ attributes); 
moving object catalog; 
variable object catalog; 
alerts catalog;
calibration metadata; 
configuration metadata; 
processing metadata; and
provenance metadata.
The science archive will consist of $\sim$2000 images
per night (for 10 years), comprising 60-100~PB
of pixel data.
This enormous LSST data archive and object database 
enables a diverse multidisciplinary research program:
astronomy \& astrophysics;
machine learning (data mining);
exploratory data analysis;
XLDB (extremely large databases); 
scientific visualization;
computational science \& distributed computing;
and inquiry-based science education (using data in the classroom).

Many possible scientific data mining use cases
are anticipated with the LSST database, including:

\begin{itemize}

\item Provide rapid probabilistic classifications for all 
10,000 LSST events each night;

\item Find new ``fundamental planes'' of 
correlated astrophysical parameters ({\it{e.g.}}, 
the fundamental plane of Elliptical galaxies) \cite{fundplane};

\item Find new correlations, associations, relationships 
of all kinds from 100+ attributes in the LSST science database,
integrated with distributed VO-accessible data;

\item Compute multi-point multi-dimensional
correlation functions over the full panoply
of astrophysical parameter spaces;

\item Discover zones of avoidance in interesting 
parameter spaces ({\it{e.g.}}, period gaps);

\item Discover new properties of known classes;

\item Discover new and improved rules for classifying known 
classes of objects ({\it{e.g.}}, photometric redshifts) \cite{way2006};

\item Discover new and exotic classes of astronomical objects;

\item Identify novel, unexpected temporal behavior 
in all classes of objects \cite{pacz2000};

\item Hypothesis testing -- verify existing (or generate new) 
astronomical hypotheses with strong statistical confidence, 
using millions of training samples;

\item Serendipity --  discover rare one-in-a-billion 
objects through novelty detection; 

\item Image processing --  identify non-astronomical features,
classify them, and separate them from the astronomical
catalog inputs \cite{salzberg95, waniak06}; 
and

\item Quality assurance --  identify system glitches,
instrument anomalies, and pipeline errors through
near-real-time deviation detection.

\end{itemize}

Some of the data mining research challenge areas posed
by the arrival of petascale scientific databases include:

\begin{itemize}

\item indexing and associative memory techniques
(trees, graphs, networks) for multi-attribute 
(highly-dimensional) astronomical 
databases (beyond RA-Dec indexing);

\item scalability of statistical,
computational, machine learning, and data mining algorithms
to multi-petabyte scales;


\item algorithms for optimization of simultaneous multi-point fitting across massive multi-dimensional data cubes;

\item multi-resolution methods
and structures for exploration of petascale databases;

\item petascale analytics for visual exploratory data analysis of massive databases; 
and

\item rapid query, search, and retrieval algorithms
for petabyte databases.

\end{itemize}

Additional and more in-depth discussion of the petascale
data challenges posed by the LSST sky survey are available
(at www.lsst.org/Project/docs/data-challenge.pdf
and universe.ucdavis.edu/docs/LSST\_petascale\_challenge.pdf).

\subsection{A Classification Broker for Astronomy}
\label{s:broker}

We envision
an astroinformatics (data-intensive astronomy) 
research paradigm (for data integration and
mining) to address the petascale 
needs of large astronomical 
surveys \cite{borne2009, borne08}. 
The impending data loads surpass 
those of the Sloan Digital Sky Survey by 1000-10,000 
times, while the time-criticality requirement 
(for event/object classification and characterization) 
drastically drops from months (or weeks) down to minutes 
(or tens of seconds). In addition to the follow-up 
classification problem (described earlier), we 
will want to find every possible new scientific 
discovery (pattern, correlation, relationship, outlier, 
new class, etc.) buried within these new enormous databases. 
This might lead to a petascale data mining compute engine 
that runs in parallel alongside the data archive -- to test
every possible 
N-point correlation, 
multi-parameter association, and
classification rule.  
In addition to such a ``batch discovery machine'', 
a rapid-response data mining engine 
({\it{i.e.}}, classification broker) is needed in order to 
produce and distribute scientifically robust near-real-time 
classifications of astronomical sources, events, objects, or 
event host objects ({\it{e.g.}}, we need
the redshift of the host galaxy in order
to interpret and classify a supernova accurately)
\cite{borne08, bloom2008, mahabal2008}. 
These classifications are derived from 
integrating and mining data, information, and knowledge 
from multiple distributed VO-accessible data repositories,
robotic telescopes, and astronomical alert networks world-wide.  
Incoming event alert data will be subjected to a suite of 
machine learning (ML) algorithms for event classification, 
outlier detection, object characterization, and novelty discovery
\cite{becker2008, borne08, bloom2008, mahabal2008, 
ball2006, ball2007b}.  
Probabilistic ML models will produce rank-ordered lists,
to guide follow-up observations on the 10-100K 
alertable astronomical events that will be identified each 
night by the LSST sky survey alone.
The classification broker will thereby enable rapid follow-up 
science for the most important and exciting astronomical 
discoveries of the coming decade, on a wide range of time 
scales from seconds to days, corresponding to a plethora 
of exotic astrophysical phenomena.


\begin{theacknowledgments}
  We thank our LSST (www.lsst.org) collaborators
for their valuable contributions.
\end{theacknowledgments}



\bibliographystyle{aipprocl} 

\end{document}